\documentstyle[epsf,12pt]{article}
\newcommand{\newsection}{    
\setcounter{equation}{0}
\section}
\def\appendix#1{
\addtocounter{section}{1}
\setcounter{equation}{0}
\renewcommand{\thesection}{\Alph{section}}
\section*{Appendix \thesection\protect\indent #1}
\addcontentsline{toc}{section}{Appendix \thesection\ \ \ #1}
}
\newcommand{\rf}[1]{(\ref{#1})}

\def\be{\begin{equation}}
\def\ee{\end{equation}}
\newcommand{\beq}{\begin{equation}}
\newcommand{\eeq}{\end{equation}}
\newcommand{\bea}{\begin{eqnarray}}
\newcommand{\eea}{\end{eqnarray}}

\newcommand{\om}{\omega}

\newcommand{\Tr}{{\,\rm Tr}\:}

\begin{document}
\topmargin 0pt
\oddsidemargin 5mm
\headheight 0pt
\headsep 0pt
\topskip 9mm
\pagestyle{empty}
\hfill DTP-97-63

\hfill SPhT-98/007

\hfill NBI-HE-97-63

\addtolength{\baselineskip}{0.20\baselineskip}
\begin{center}
\vspace{26pt}
{\large \bf Hamiltonian Cycles on a Random Three-coordinate Lattice}


\vspace{26pt}

{\sl B.\ Eynard}\hspace*{0.05cm}\footnote{
E-mail: 
Bertrand.Eynard@durham.ac.uk,
\\Permanent address: 
Service de Physique Th\'{e}orique de Saclay, 
F-91191 Gif-sur-Yvette Cedex}\\
\vspace{6pt}
Department of Mathematical Sciences \\
University of Durham, Science Labs. South Road \\
Durham DH1 3LE, UK\\

\vspace{18pt}
{\sl E.\ Guitter}\hspace{0.025cm}\footnote{E-mail:
guitter@spht.saclay.cea.fr} \\
Service de Physique Th\'{e}orique de Saclay \\
F-91191 Gif-sur-Yvette Cedex, France \\

\vspace{18pt}
{\sl C. Kristjansen}\hspace{0.025cm}\footnote{E-mail:kristjan@nbi.dk}
\\ 
\vspace{6pt}
The Niels Bohr Institute \\
 Blegdamsvej 17,
DK-2100 Copenhagen \O, Denmark \\
\end{center}
\vspace{20pt} 
\begin{center}
Abstract
\end{center}
Consider a random three-coordinate lattice of spherical topology having
$2v$ vertices and being densely 
covered by a single closed, self-avoiding walk, i.e.\
being equipped with a Hamiltonian cycle.
 We determine the number of such objects as a function of $v$. Furthermore we 
express the partition function of the corresponding statistical model as an 
elliptic integral.

\vfill{\noindent PACS codes: 05.20.y, 04.60.Nc, 02.10.Eb }
\newpage

\newsection{Introduction}
\pagestyle{plain}
\setcounter{page}{1}
Hamiltonian cycles play an important role in the study of the configurational
statistics of polymers~\cite{polymers}.
Given a lattice, a Hamiltonian cycle is defined as a closed, self-avoiding
walk which visits each vertex of the lattice once and only once.
Typically, in polymer physics, one has been interested in calculating the
number of Hamiltonian cycles for a lattice with a certain regular structure
and a given number of vertices.  This counting problem has been exactly solved 
only in a few cases, namely for the two-dimensional Manhattan oriented square
lattice~\cite{Kas63}, the two-dimensional ice lattice~\cite{Lie67} and the
two-dimensional hexagonal lattice~\cite{Suz88,BSY94}. A field theoretical
approach to the problem of counting Hamiltonian cycles on regular lattices
was invented by H.\ Orland et.\ al.~\cite{OID85} and has recently been
further developed by S.\ Higuchi~\cite{Hig97}.

Here we address the problem of counting Hamiltonian cycles on random lattices.
Whereas on a regular lattice the choice of boundary conditions has to
be carefully considered and plays a crucial role for the outcome of
the analysis (see for instance~\cite{polymers} for a discussion) on a random
lattice no such complications occur. Furthermore one might speculate
if the complex dynamics of polymers does not call for a random lattice
rather than a regular one.

Consider a random three-coordinate lattice of spherical topology having 
$2v$ vertices and being equipped with a Hamiltonian cycle. In the present
paper we aim at determining the number of such objects as a function of $v$.
The objects in question can also be viewed as spherical triangulations
consisting of $2v$ triangles and being densely covered by a single
closed, self-avoiding walk (where it is understood that the walk
consists of links belonging to the dual lattice). Not any
three-coordinate graph has a Hamiltonian cycle. Obviously,
one-particle-reducible graphs do not, but there also exist
one-particle-irreducible graphs which can not be equipped with a
Hamiltonian cycle.  We show an example of such a graph along with the
corresponding triangulation in figure~1.
\begin{figure}[htb]
\centerline{\epsfxsize=14.truecm\epsfbox{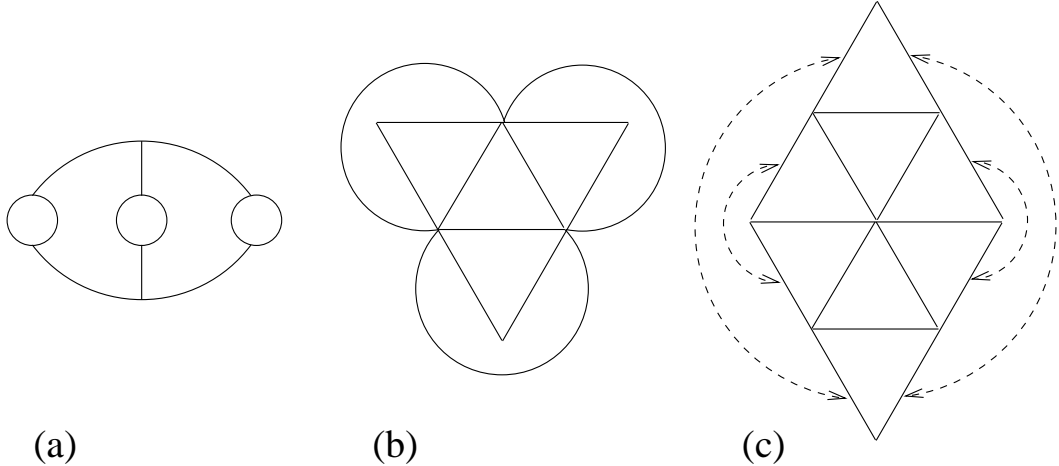}}
\caption{A three-coordinate one-particle-irreducible graph, which does
not have a Hamiltonian cycle (a), and the corresponding triangulation
(b) or equivalently (c).}
\end{figure}
In order to study the above mentioned combinatorial problem we consider a
matrix model which generates graphs or triangulations of the relevant
type, namely the $n\rightarrow 0$ limit of the fully packed $O(n)$
model on a random lattice. The $O(n)$ model is a so-called loop-gas
model and the loops, being closed, self-avoiding and non-intersecting
play the role of the walks. In the fully packed model all vertices of
the generated graphs are visited by a loop and in the limit
$n\rightarrow 0$ only one loop remains. The regular lattice version of
the fully packed $O(n)$ model  was studied in 
references~\cite{Suz88,BN94,BSY94,Kas94}.

The $O(n)$ model on a random lattice comes in two different versions,
an unrestricted one~\cite{Kos89,EK96} and a restricted one~\cite{EK97}. In
the latter the loops are restricted to having even length while in the
former there is no restriction on the loop length. On a regular
lattice all loops necessarily have even length. The two different
versions of the random lattice $O(n)$ model have different critical
properties. The restricted $O(n)$ model belongs to the universality
class of the unrestricted $O(\frac{n}{2})$ model~\cite{EK97}. 
The fully packed, restricted $O(n)$ model with $n=2$ is of particular
interest because it describes the generalisation of Baxter's
three-colour problem~\cite{Bax70} to a random
lattice~\cite{Three,EK97}. As long as one considers triangulations
consisting only of an even number of triangles (such as those
corresponding to a two-dimensional manifold without boundaries) the
$n\rightarrow 0$ limit of the fully packed, unrestricted $O(n)$ model 
coincides with
the $n\rightarrow 0$ limit of the fully packed, 
restricted $O(n)$ model. Since the
unrestricted $O(n)$ model on a random lattice is mathematically easier
to handle than the restricted one, we shall take as our starting point
the unrestricted one. We note, however, that our $n\rightarrow 0$
solution applies to the restricted model as well.

\newsection{The Model}
Our starting point is the following matrix model
\beq
Z(g)=\int_{N\times N}dA\prod_{i=1}^n dB_i 
     \exp\left\{ -N\Tr \left[ \frac{1}{2}A^2+
\frac{1}{2}\sum_{i=1}^nB_i^2
-\sqrt{g}\sum_{i=1}^nAB_i^2\right]\right\}
\label{M1}
\eeq
where $A$ and $B_i$ are hermitian $N\times N$ matrices.
This matrix model generates triangulations which are densely covered
by closed, self-avoiding and non-intersecting loops which come in
$n$ different colours. More precisely we can
write the free energy $F(g)=\frac{1}{N^2}\log Z(g)$ as
\bea
F(g)&=&\sum_{h=0}^{\infty}N^{-2h}F_h(g), \nonumber \\
F_h(g)&=&\sum_{v=1}^{\infty} {\cal N}_h(2v)\, g^v,\hspace{1.0cm}
{\cal N}_h(2v)=\sum_{i=0}^v n^i {\cal N}_h^{(i)}(2v).
\eea
The quantity ${\cal N}_h^{(i)}(2v)$ counts the number of genus $h$
triangulations built from $2v$ triangles and densely covered by 
$i$ closed, self-avoiding and non-intersecting
loops$\,$\footnote{Here it is
 understood that a triangulation, $T$, with
a loop configuration $\{ {\cal L}\}$ is counted with the weight
$1/\mbox{Aut}(T_{\{ {\cal L } \}})$ where $\mbox{Aut}(T_{\{ {\cal L} \}})$ is
the order of the auto-morphism group of the triangulation $T$ with
the loop configuration $\{ {\cal L} \}$}.
 In the following we shall aim at determining ${\cal N}_0^{(1)}(2v)$, 
i.e.\ the number of  spherical
triangulations consisting of $2v$ triangles and being  
densely covered by a {\it single} closed, self-avoiding walk.
\noindent
First, let us write instead of~\rf{M1}
\bea
{\cal Z}(g)&=&\left(\sqrt{g}\right)^{N^2} Z(g)  \nonumber \\
&=&\int_{N\times N}dA
\prod_{i=1}^ndB_i \exp \left\{-N\Tr\left[
\frac{1}{2g}A^2+\frac{1}{2}\sum_{i=1}^nB_i^2
-\sum_{i=1}^n AB_i^2 \right]\right\}.
\label{M2}
\eea
Furthermore, let us introduce a resolvant $W(z)$ corresponding to
the model given by the partition function ${\cal Z}(g)$\
\beq
W(z)=\langle \frac{1}{N}\Tr \frac{1}{z-A}\rangle
=\sum_{h=0}^{\infty}N^{-2h}\,W_h(z).
\eeq
We shall be interested only in the large-$N$ limit of the resolvant, 
$W_0(z)$
\beq
W_0(z){\mathop\sim_{z\to \infty}^{}} \,\,\,\,\sum_{m=0 }^\infty
\frac{t_m}{z^{m+1}},\hspace{0.7cm} 
t_m=\langle \frac{1}{N}\Tr A^{m}\rangle_{h=0}.
\eeq
Let us consider the quantity $t_2$
\bea
t_2&=&\frac{2g^2}{N^2}\frac{d}{dg}\log\left\{
(\sqrt{g})^{N^2}Z(g)\right\}_{h=0}
   =g+2g^2\frac{d}{dg}F_0(g) \nonumber \\
   &=&g+2\sum_{v=0}^{\infty}\sum_{i=0}^{v}n^i {\cal N}_0^{(i)}(2v)\,v\,g^{v+1}.
\eea
We  see that we can achieve our aim of calculating
${\cal N}_0^{(1)}(2v)$ by determining the contribution to $t_2$ 
which depends
linearly on $n$. In order to do so, we shall make use of the well-known
loop equation for the resolvant~\cite{Kos89,EK96}. This equation can be
derived by making use of the invariance of the matrix integral~\rf{M2}
under analytic redefinitions of the fields and reads in the large-$N$
(spherical) limit
\beq
\frac{1}{g}\left(zW(z)+(1-z)W(1-z)-2\right)=\left(W(z)\right)^2
+\left(W(1-z)\right)^2+nW(z)W(1-z) \label{loop}
\eeq
We now introduce the following functions
\bea
\xi_m(z)&=&\frac{1}{z^m}+\frac{1}{(1-z)^m}, \nonumber \\
\xi_{m,p}(z)&=&\frac{1}{z^m(1-z)^p}+\frac{1}{(1-z)^m z^p}.
\eea
The functions $\xi_{m,p}(z)$ can be expressed in terms of the $\xi_k(z)$ as
follows 
\beq
\xi_{m,p}(z)=\sum_{k=1}^m \left( \begin{array}{c} m+p-1-k \\ p-1\end{array}
\right)\xi_k(z)
+\sum_{k=1}^p\left( \begin{array}{c} m+p-1-k \\ m-1 \end{array}\right)
\xi_k(z)
\label{decompose}
\eeq
and in terms of the $\xi$-functions we can write the loop equation~\rf{loop} as
\beq
\frac{1}{g}\sum_{l=1}^{\infty}t_l\xi_l =
\sum_{k,l}t_k t_l \xi_{k+l+2}+ \frac{n}{2}
\sum_{k,l=0}^{\infty}\xi_{k+1,l+1} t_k t_l.
\label{before}
\eeq
Using the relation~\rf{decompose} and identifying the coefficients of each
$\xi_m$ in~\rf{before} we get for $m\geq 2$
\beq
\frac{1}{g}t_m=\sum_{k=0}^{m-2}t_k t_{m-k-2} + 
n\sum_{l,k=0}^{\infty}t_{m+k-1}t_l 
\left(\begin{array}{c} k+l \\ l\end{array}\right). 
\eeq
This equation can be solved perturbatively in $n$ and $g$. 
First, let us set $t_m=g^{m/2}T_m$. Then we have
\beq
T_m=\sum_{k=0}^{m-2}T_k T_{m-k-2}+
n\sum_{v=1}^{\infty}\sum_{k=1}^v 
\left(\begin{array}{c} v-1 \\ k-1 \end{array}\right) T_{v-k}T_{m+k-2}\,g^{v/2}
\label{iter}
\eeq
{}From this we see that the $T_k$'s have the following expansion 
\beq
T_k=\sum_{v=0}^\infty\sum_{i=0}^{v} n^i\, T_k^{(i,v)} g^{v/2}
\eeq
and that 
\beq
2v\, {\cal N}_0^{(1)}(2v)=\, T_2^{(1,2v)}.
\eeq

\newsection{Exact solution as $n\rightarrow 0$}
For $n=0$ our model~\rf{M2} is nothing but the Gaussian one-matrix
model and we have
\beq
T_{2k}^{(0,0)}=c_k\equiv \frac{(2k)!}{k!(k+1)!},
\hspace{0.7cm}T_{2k+1}^{(0,0)}=0.
\label{Catalan}
\eeq
Here $c_k$ is the Catalan number of order $k$ which counts in particular 
the number of distinct configurations of $k$ non-intersecting arches drawn 
on top of a line.

Equating the terms linear in $n$ in equation~\rf{iter} we see that
\bea
T_2^{(1,2v)}&=&2v\,{\cal N}_0^{(1)}(2v)  
= \sum_{k=1}^v 
\left( \begin{array}{c} 2v-1 \\ 2k-1\end{array} \right) T_{2k}^{(0,0)}\, 
T_{2v-2k}^{(0,0)} \nonumber \\
&=& \frac{2(2v-1)!}{(v+2)!(v-1)!}
\sum_{k=1}^v \left(\begin{array}{c} v+2\\ k+1 \end{array}\right)
\left(\begin{array}{c} v-1 \\ k-1 \end{array}\right) \nonumber \\
&=&\frac{2(2v-1)!(2v+1)!}{(v+2)!(v+1)!\,v!\,(v-1)!}. 
\label{result}
\eea
To prove the last equality one can for instance compare the terms
of power zero in $x$ on the two sides of the obvious identity
\beq
\frac{1}{x}\,\left(1+x\right)^{v+2}\,
\frac{1}{x}\left(1+\frac{1}{x}\right)^{v-1}=
\frac{1}{x^{v+1}}\left(1+x\right)^{2v+1}. 
\eeq 
Next, let us define a partition function for spherical triangulations 
densely covered by a single closed and self-avoiding walk
\bea
F_0^{(1)}(g)&\equiv&
\sum_{v=1}^\infty {\cal N}_0^{(1)}(2v)\, g^v \nonumber \\
    &=&\sum_{v=1}^\infty \frac{(2v-1)!(2v+1)!}{(v+2)!(v+1)!(v!)^2}\, g^v 
\nonumber \\
   &=& \frac{1}{2g^2}\sum_{v=1}^\infty \frac{(2v)! (2v+1)!}{(v!)^3 (v+1)!} \,
\frac{1}{(v+2)(v+1)\,v}\, g^{v+2}.
\eea
Here $g$ plays the role of a cosmological constant. Using~\cite{Han75}
we find
\bea
\frac{\partial^3}{\partial g^3}\left(2g^2 F_0^{(1)}(g)\right)&=&
\frac{1}{g}\sum_{v=1}^\infty \frac{(2v)!(2v+1)!}{(v!)^3 (v+1)!}\, g^v 
\nonumber \\
&=&\frac{1}{4\pi g^2}\left[ K(4\sqrt{g})-E(4\sqrt{g})\right]
-\frac{1}{g}
\hspace{0.7cm}\mbox{for}\hspace{0.7cm} 0<g<\frac{1}{16}
\label{fenergy}
\eea
where $K$ and $E$ are the complete 
elliptic integrals of the first and second kind
respectively.
The radius of convergence of the series above is
$g=g_c=\frac{1}{16}$. 
Using the analyticity properties of the elliptic integrals one gets
\beq
F_0^{(1)}(g)\sim \frac{1}{4\pi g_c^2}\, (g_c-g)^3\, \log (g_c-g) 
\hspace{0.5cm}\mbox{for}\hspace{0.5cm} g\sim g_c.
\eeq
This shows that the value of the critical index $\gamma$ defined by
$F_0^{(1)}(g)\sim (g_c-g)^{2-\gamma}$ takes the value $-1$ which is the
value characteristic of $c=-2$ conformal matter coupled to quantum
gravity. This scaling is in accordance with the scaling obtained 
in~\cite{DK90}.

\newsection{Direct combinatorial approach}
The number ${\cal N}_0^{(1)}(2v)$ can be derived more directly from 
the following purely combinatorial argument. For each spherical 
three-coordinate lattice covered by a Hamiltonian cycle, let us 
choose one particular link {\it visited} by the cycle. For lattices 
with $2v$ vertices, there are exactly $2v$ such visited links. By 
cutting the link and pulling apart the two created end-points, we can deform 
continuously on the sphere the (now open) Hamiltonian cycle into a 
straight line. Since the path is Hamiltonian, all the vertices of the original 
lattice lie on this line. To fully re-construct the lattice 
configuration, we simply need to specify the image under the deformation 
of all the {\it unvisited} links. Since we are dealing with a three-coordinate 
lattice, each vertex on the line is the extremity of exactly one such 
unvisited link. After the deformation, unvisited links thus form a
system of $v$ non-intersecting arches which connect the $2v$ vertices in
pairs, see figure~2.
\begin{figure}[hbt]
\vspace{0.5cm}
\centerline{\epsfxsize=15.truecm\epsfbox{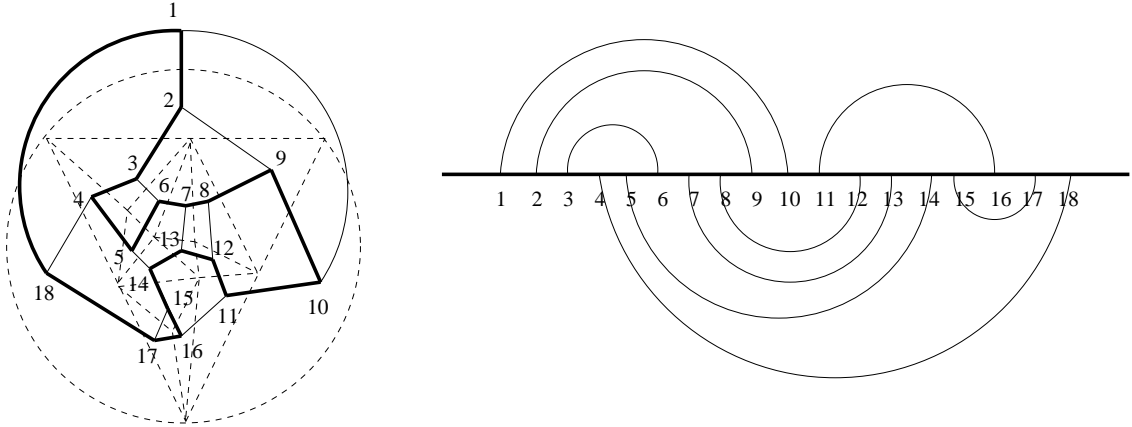}}
\vspace{0.5cm}
\caption[ppp]{Left: A triangulation consisting of 18 triangles, equipped with a
Hamiltonian cycle. Right: The corresponding systems of arches.}
\end{figure} 
Pairs can be connected either above or below the straight line. 
Note that by choosing which side of the line is the top or
the bottom side, we artificially double the number of configurations.

Under the above deformation, the same arch configuration is obtained
exactly $\mbox{Aut}(T_{\{ {\cal L } \}})$ times, where 
$\mbox{Aut}(T_{\{ {\cal L} \}})$ is the order of the auto-morphism group 
of the original triangulation $T$ with the loop configuration $\{ {\cal L} \}$.
This factor cancels out the pre-factor $1/\mbox{Aut}(T_{\{ {\cal L } \}})$ 
in the definition of ${\cal N}_0^{(1)}(2v)$. All distinct 
arch configurations are thus counted exactly once.
We can finally identify $2\times 2v \times {\cal N}_0^{(1)}(2v)$ as
the number of distinct systems of $v$ non-intersecting arches
connecting $2v$ vertices on a straight oriented line from above and/or
below this line.

By decomposing the $v$ arches into a system of $k$ arches above
the line and $v-k$ arches below, we get
\beq
4v\,{\cal N}_0^{(1)}(2v)=\sum_{k=0}^v \left(\begin{array}{c} 2v \\ 2k
\end{array}\right)  c_k \, c_{v-k},
\eeq
where the binomial coefficient comes from the specification of the vertices 
which are connected from, say, above and where $c_k$ denotes the number of 
distinct 
configurations of the $k$ arches above the line. 
This number is nothing but the Catalan number given in \rf{Catalan}, 
which leads to 
\bea
4v\,{\cal N}_0^{(1)}(2v) &=&  \sum_{k=0}^v
\left( \begin{array}{c} 2v \\ 2k\end{array} \right) c_k\,
c_{v-k} \nonumber \\
&=&\frac{(2v)!(2v+2)!}{(v+2)!((v+1)!)^2\,v!}=
c_v\,c_{v+1}
\eea
in agreement with \rf{result}. 
\newsection{Conclusion}
We have solved the non-trivial combinatorial problem of determining
the number of spherical triangulations consisting of $2v$ triangles
and being densely covered by a single self-avoiding and closed walk. 
Let us define the entropy exponent of such objects $\om_H$, as
\beq
\log \om_H=\lim_{v\rightarrow \infty} \frac{1}{2v} 
\log \left({\cal N}_0^{(1)}(2v)\right).
\eeq
{}From~\rf{fenergy} we see that
\beq
\om_H=4.
\eeq
For triangulations without decorations the corresponding exponent
takes the value~\cite{BIPZ78}
\beq
\om_H^{(T)}=2\cdot 3^{3/4}
\eeq
and for triangulations corresponding to one-particle-irreducible
three-coordinate graphs
\beq
\om_H^{(1PI)}=\frac{16}{3\sqrt{3}}.
\eeq
As mentioned in the introduction, one-particle-reducible graphs do not
contribute to ${\cal N}_0^{(1)}(2v)$ since they do not have any
Hamiltonian cycles. Not any one-particle-irreducible graph contributes
to ${\cal N}_0^{(1)}(2v)$ either, but those which do are counted with a
weight which equals the number of ways they can  be equipped by a 
Hamiltonian cycle. 

We call the attention of the reader to the fact that
we have obtained an exact expression for the spherical contribution to
the free energy of the restricted as well as the unrestricted, fully 
packed $O(n)$ model on a random lattice in the limit $n\rightarrow 0$.
An exact solution of the complete, restricted $O(n)$ model is still
lacking. Such a solution would amongst others provide
us with an exact solution of the three-colour problem on a random
lattice~\cite{Three, EK97}.
The results of the present paper might be viewed as a 
step towards the exact solution of this problem.
\begin{figure}[htb]
\centerline{\epsfxsize=14.truecm\epsfbox{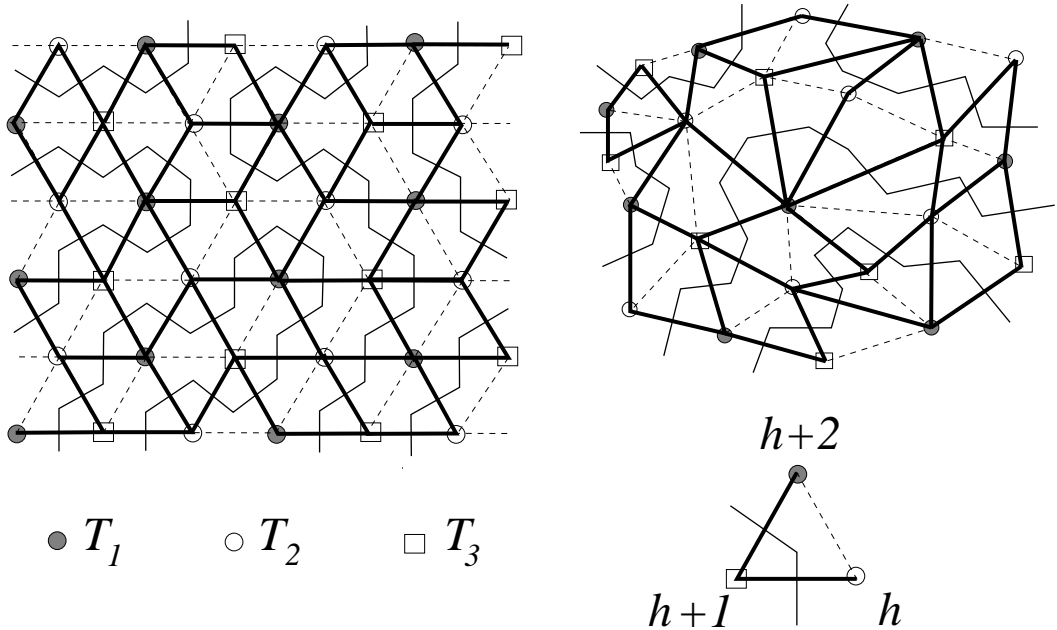}}
\caption{Construction of the SOS degree of freedom from the fully packed
loops, on a regular triangular lattice and on a random vertex three-colourable
triangulation. The thick lines indicate links whose dual link is occupied
by a loop. The height variable $h$, defined on the vertices of the lattice,
is such that the difference of height is $\pm 1$ for nearest neighbours
connected by a thick line and $\pm 2$ for nearest neighbours connected
by a dashed line. The sign is fixed by demanding that $h$ and 
the number $i$ of the corresponding sub-lattice $T_i$ are equal modulo 3. 
For the regular lattice, this SOS height variable corresponds precisely to
the underlying picture of a 2d piling of elementary cubes.}
\end{figure}
To finish, we may ask how important is the Hamiltonian nature
of the walks, or more generally whether or not the {\it fully packing}
of the (restricted or unrestricted) $O(n)$ model is a relevant
constraint. For an $O(n)$ model on the regular honeycomb lattice, it 
was explained in~\cite{BN94} that the fully packed fixed point differs from 
the low temperature dense phase fixed point by the emergence of 
an additional SOS degree of freedom which can be constructed if and 
only if the loops are fully packed. This degree of freedom is defined 
as a height variable $h_v$ living on the vertices $v$ of the dual 
triangular lattice (see figure~3). Indeed, this lattice is naturally 
divided into 
three alternating sub-lattices $T_{i}$, $i=1,2,3$. Fully packed loops 
can then be translated into height variables by demanding that 
$h_v=i \ {\rm mod}\ 3$ for $v\in T_{i}$ and by imposing for nearest 
neighbours $v$ and $v^{\prime}$ that $|h_v-h_{v^\prime}|=1$ if the link 
between $v$ and $v^\prime$ has its dual link occupied by a loop and 
$|h_v-h_{v^\prime}|=2$ if not. This extra SOS degree of freedom results 
in a shift by 1 of the central charge of the fully packed fixed point 
with respect to the usual fixed point for the dense phase. On a random 
lattice, vacancies (i.e. unvisited vertices) can be introduced for 
instance by adding a term proportional to $A^3$ in the potential 
in~\rf{M1}. As shown in~\cite{EK96}, on a random lattice
the nature of the dense phase
fixed point is not changed by eliminating vacancies
and the constraint of fully packing is thus irrelevant in this sense.
This is also what we find here for $n\rightarrow 0$ where we get
$c=-2$ for Hamiltonian walks on a random lattice, as 
for the dense phase of the $O(n\rightarrow 0)$ model and not $c=-1$ as for 
the fully packed $O(n\rightarrow 0)$ fixed point describing Hamiltonian walks 
on the honeycomb lattice.
This result is not surprising since, on a random lattice, fully packed 
loops cannot be transformed anymore into an SOS height variable and we 
thus expect no shift in the central charge. As a natural question, we 
may ask whether it is possible to preserve the SOS variable on a random 
triangular lattice. A SOS variable can be defined in the same way as 
above if the random lattice can be divided into three alternating 
sub-lattices. In other words, one needs that the vertices of the lattice 
can be coloured in three colours such that nearest neighbours have 
different colours or, stated in yet another way, that any given vertex
is shared by an even number of triangles.
A vertex three-colouring problem of the above mentioned type
was discussed and solved in~\cite{DFEG97}. 
Its coupling to a fully packed $O(n)$ model is still an
open question but, according to the above discussion we expect the same
shift in the central charge as for the regular model. 
Note that fully packed loops automatically have even length on a vertex 
three-colourable triangular lattice. The case $n=2$ is of particular 
interest since it would describe
a problem of three-colouring of both the vertices and the edges of random 
triangulations, a problem equivalent to the folding of random triangulations, 
as explained in~\cite{EK97,DFEG97}. 
Based on the discussion above, we expect $c=2$ for this model, as for the 
fully packed $O(2)$ model on a regular lattice~\cite{BN94}.

\vspace{15pt}
\noindent
{\bf Acknowledgements} \hspace{0.3cm} 
B.\ Eynard acknowledges the support and the hospitality of the Niels
Bohr Institute as well as the support of the European Union through their 
TMR programme, contract No.\ ERBFMRXCT 960012. We thank F. David for
a critical reading of the manuscript.

\end{document}